\documentclass{article}

\usepackage{graphics}
  
\newcommand{\prl}[3]{{Phys. Rev. Lett.}{\bf #1}{(#2)}{#3}}
\newcommand{\prd}[3]{{Phys. Rev.}{\bf D#1}{(#2)}{#3}}
\newcommand{\pr}[3]{{Phys. Rev.}{\bf #1}{(#2)}{#3}}
\newcommand{\hepph}[1]{hep-ph/#1}
\newcommand{\hepex}[1]{hep-ex/#1}
\newcommand{\arnps}[3]{{Ann. Rev. Nucl. Part. Sci.}{\bf #1}{(#2)}{#3}}
\newcommand{\plb}[3]{{Phys. Lett.}{\bf B#1}{(#2)}{#3}}
\newcommand{\npb}[3]{{Nucl. Phys.}{\bf B#1}{(#2)}{#3}}
\newcommand{\mpla}[3]{{Mod. Phys. Lett.}{\bf A#1}{(#2)}{#3}}

\newcommand{\gae}{\stackrel{>}{\sim}}

\begin{titlepage}

\title{Thinking About Top: \\ Looking Outside The Standard Model \footnote{Talk
    given at 
    the Thinkshop on Top Quark Physics at Run II, 16-18 October, 1998,
    Fermilab, Batavia, IL}}

\author{Elizabeth H. Simmons\thanks{e-mail address: simmons@bu.edu}\\
Department of Physics, Boston University\\
590 Commonwealth Avenue, Boston, MA  02215}

\begin{document}

\maketitle
\bigskip
\begin{picture}(0,0)(0,0)
\put(295,250){BUHEP-99-17}
\put(295,235){hep-ph/9908511}
\end{picture}
\vspace{24pt}
 
\begin{abstract}
The top quark is by far the heaviest known fermion \cite{ehs-disc-top}.  In
consequence, experiment is just beginning to explore its properties, and some
of them may 
yet prove to be distinctly non-standard.  The very size of the top quark's
mass even hints at the possibility of a special role for top in electroweak
symmetry breaking.  This talk examines the top quark in the context of
physics beyond the standard model, and discusses how Run II can help
elucidate the true nature of top.

 \pagestyle{empty}
\end{abstract}
\end{titlepage}
 
\null\vspace{-.5cm}
\section{Introduction}
\setcounter{equation}{0} 

This talk looks beyond the role and properties of the top quark within
the standard model.  Experiments at Run II at the Fermilab Tevatron will
certainly have much of interest to say about the top quark even considered
strictly as a standard model particle (see e.g. the preceding Thinkshop talk
by Scott Willenbrock \cite{ehs-willen}).  But if physics beyond the standard
model exists at energy  
scales accessible to upcoming experiments, the possibilities for top quark
physics are greatly expanded.  We will start by reviewing why one needs to
look beyond the standard model, then discuss specific possibilities of new
physics that could be associated with top, and consider how signs of such new
physics could manifest itself in experiment.

\section{Beyond the Standard Model}
\setcounter{equation}{0}

Two central concerns of particle theory at the close of the millenium are
finding the cause of electroweak symmetry breaking and identifying the origin
of flavor symmetry breaking by which the quarks and leptons obtain their
diverse masses.  The Standard Model of particle physics, based on the gauge
group $SU(3)_c \times SU(2)_W \times U(1)_Y$ accommodates both symmetry
breakings by including a fundamental weak doublet of scalar (``Higgs'')
bosons ${\phi = {\phi^+ \choose \phi^0}}$ with potential function $V(\phi) =
\lambda \left({\phi^\dagger \phi - \frac12 v^2}\right)^2$.  However the
Standard Model provides no explanation of the dynamics responsible for the
generation of mass.  
Furthermore, the scalar sector suffers from
\begin{figure}[ht]
{\scalebox{.36}{\includegraphics[-5.5in,0in][3in,1.25in]{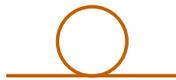}}
\caption{$ M_H^2\ \propto\ \Lambda^2$}\label{fig2}}
\end{figure}
\noindent 
two serious problems.  The scalar mass is unnaturally sensitive to the
presence of physics at any higher scale $\Lambda$ (e.g. the Planck scale), as
shown in figure \ref{fig2}.  This is known as the gauge hierarchy problem.
In addition, if the scalar must provide a good description of physics up to
arbitrarily high scale (i.e., be fundamental), the 
\begin{figure}[ht]
{\scalebox{.4}{\includegraphics[-5in,0in][3in,1.25in]{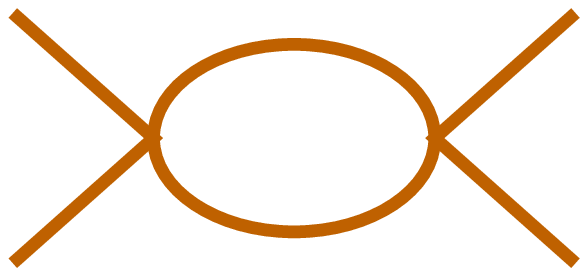}}
{\caption{$\beta(\lambda)\ = \ {{3\lambda^2}\over{2\pi^2}}\ > \
  0$}}\label{fig3}}
 \end{figure} 
\noindent scalar's self-coupling
($\lambda$) is driven to zero at finite energy scales as indicated in figure
\ref{fig3}.  That is, the scalar field theory is free (or ``trivial''). Then
the scalar cannot fill its intended role: if $\lambda = 0$, the electroweak
symmetry is not spontaneously broken.  The scalars involved in electroweak
symmetry breaking must therefore be composite at some finite energy scale.
We must seek the origin of mass in physics that lies beyond the standard
model with its fundamental scalar doublet.  

One interesting possibility (denoted ``dynamical electroweak symmetry
breaking'' \cite{ehs-1-dynam}) is that the compositeness of the scalar states
involved in electroweak symmetry breaking could manifest itself at scales not
much above the electroweak scale $v \sim 250 GeV$.  In these theories, a new
strong gauge interaction with $\beta < 0$ (e.g technicolor) breaks the chiral
symmetries of a set of massless fermions $f$ at a scale $\Lambda \sim 1$TeV.
If the fermions carry appropriate electroweak quantum numbers, the resulting
condensate $\langle \bar f_L f_R \rangle \neq 0$ breaks the electroweak
symmetry as desired.  The logarithmic running of the strong gauge coupling
renders the low value of the electroweak scale (i.e.  the gauge hierarchy)
natural.  The absence of fundamental scalar bosons obviates concerns about
triviality.

\begin{figure}[hb] 
{\scalebox{.5}{\includegraphics[-2in,0in][3in,1.5in]{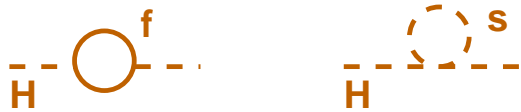}}
\caption{$\delta
      M_H^2 \sim {g_f^2\over{4\pi^2}} (m_f^2 - m_s^2) + m_s^2 log
      \Lambda^2$}\label{fig4}}
\end{figure}

Another intriguing idea is to modify the Standard Model by introducing
supersymmetry \cite{ehs-1-susy}.  The gauge structure of the minimal
supersymmetric version of the Standard Model (MSSM) is identical to that of
the Standard Model, but each ordinary fermion (boson) is paired with a new
boson (fermion) called its ``superpartner'' and two Higgs doublets are needed
to provide mass to all the ordinary fermions.  As sketched in figure
\ref{fig4}, each loop of ordinary
particles contributing to the Higgs boson's mass is now countered by a loop 
of superpartners.   
\noindent If the masses of the ordinary particles and superpartners
are close enough, the gauge hierarchy can be stabilized \cite{ehs-1-stab}.
In addition, supersymmetry relates the scalar self-coupling to gauge
couplings, so that triviality is not a concern.

Once we are willing to consider physics outside the Standard Model, 
the next question is how to find it in experiment.  One logical place to look
is in the properties of the most recently discovered state, the top 
quark. The fact that its mass is of order the electroweak scale suggests that
the top quark may afford us insight about existing non-standard models of
electroweak physics and could even play a special role in electroweak and
flavor symmetry breaking.  Since the sample of top quarks available for study
in Run I at the Tevatron was relatively small, many of the top quark's
properties are still only loosely constrained.  This opens the possibility
that the top quark could have properties that set it apart from the other
quarks.  Examples include: light related states, low-scale compositeness,
unusual gauge couplings, and a unique role in electroweak dynamics.
Fortunately, the upcoming experiments at Run II will help us evaluate these
ideas; for instance, a list of ``symptoms of new physics'' to look for
in the Run II top-pair sample is on the Thinkshop web site at \cite{ehs-demina}.

\section{Light Related States}
\setcounter{equation}{0} 

In many theories beyond the standard model, the spectrum of particles
accessible to upcoming experiments includes new states related to the top
quark. Some couple to the top quark, allowing the possibility of new
production or decay modes.  Others mix with the top quark, altering the
properties of the lighter ``top'' eigenstate we have seen relative to
standard model predictions.  As a side issue, it is worth mentioning that if
the large mass of the top arises from flavor non-universal couplings between
the top quark and new boson states, then flavor-changing neutral current
decays ($t \to c + X, u + X$) may result.

\subsection{Light Top Squarks}

\begin{figure}[ht]
{\scalebox{.55}{\includegraphics[0in,1.75in][6in,7in]{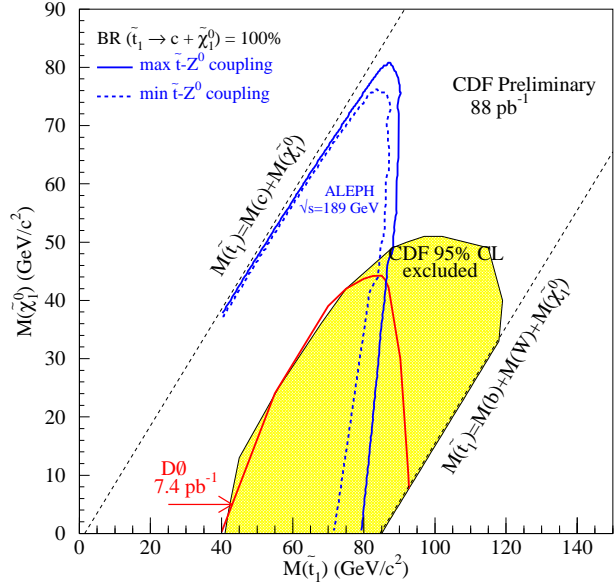}}
\caption{Searches for scalar top \protect\cite{ehs-3-slite} have excluded
  regions below the curves as shown, but still allow the stop to be lighter
  than the top .}\label{fig5}}
\end{figure}

Since supersymmetric models include a bosonic partner for each standard model
fermion, there is a pair of scalar top squarks affiliated with 
top  (one associated with $t_L$ and one, with $t_R$).  A glance at the
mass-squared matrix for the supersymmetric partners of the top quark:
\begin{center}
\begin{math}\null\hspace{-1cm}\tilde{m}_t ^2 = 
\pmatrix{\tilde{M}^2_Q + m_t^2 &\ & 
{m_t}(A_t + \mu\cot\beta)\cr 
+ M_Z^2(\frac12 - \frac23 \sin^2\theta_W)\cos2\beta &\ &\cr
\ &\ &\ \cr  {m_t}(A_t +
    \mu\cot\beta)& &\tilde{M}^2_U + m_t^2 \cr
&\ & + \frac23 M_Z^2 \sin^2\theta_W
    \cos2\beta\cr} 
\end{math}
\end{center}
reveals that the off-diagonal entries are proportional to $m_t$.  Hence, a
large top quark mass can drive one of the top squark mass eigenstates to be
relatively light.  Experiment still allows this possibility
\cite{ehs-3-slite}, as may be seen in figure \ref{fig5}.

This raises the possibility that perhaps some of the ``top'' sample observed
in Run I included top squarks\cite{ehs-3-stops}.  If the top squark is not
much heavier than 
the top quark, it is possible that $\tilde{t}\tilde{t}$ production occurred
in Run I, with the top squarks subsequently decaying to top plus neutralino
or gluino (depending on the masses of the ``inos'').  On the other hand, if
the top is a bit heavier than the stop, some top quarks produced in
$t\bar{t}$ pairs in Run I may have decayed to top squarks via $t \to
\tilde{t} \tilde{N}$ with the top squarks' subsequent decay being either
semi-leptonic $\tilde{t} \to b \ell \tilde\nu$ or flavor-changing $\tilde t
\to c \tilde{N}, c \tilde{g}$.  With either ordering of mass, it is possible
that gluino pair production occurred, followed by $\tilde{g} \to t
\tilde{t}$.

Such ideas can be tested by
studying the absolute cross-section, leptonic decays, and kinematic
distributions of the top quark events \cite{ehs-demina}.  For example, stop
or gluino production could increase the apparent $t\bar t$ production rate
above that of the standard model.  Or final states including like-sign
dileptons could result from gluino decays.

\subsection{Exotic quarks}

A variety of models propose the existence of a new charge 2/3 quark which
mixes with the top quark and alters the properties of the ``top'' state we
see from those predicted in the standard model.  In some models, the result
of the mixing is two nearly-degenerate states, which would imply that the top
sample at Run I contained an admixture of exotic quarks.
The larger top sample in Run II could make this apparent.  In other models,
the mass matrix of the top and its exotic partner is of a seesaw form
\begin{center}
\begin{math}
\pmatrix{\bar{t_L} & \bar{t}_L\prime\cr}
\pmatrix{0 & m_1 \cr m_2 & M \cr}
\pmatrix{t_R \cr t_R\prime \cr}
\end{math}
\end{center}
so that the extra state can be considerably heavier than the observed top
quark \cite{ehs-mix-lite}.  In this case, the best clue to the presence of
new physics might be alterations in the branching fractions of top quark
decays.

\subsection{Charged scalar bosons}

\begin{figure}[ht]
{\scalebox{.45}{\includegraphics[-1.75in,1.5in][6.5in,8.5in]
    {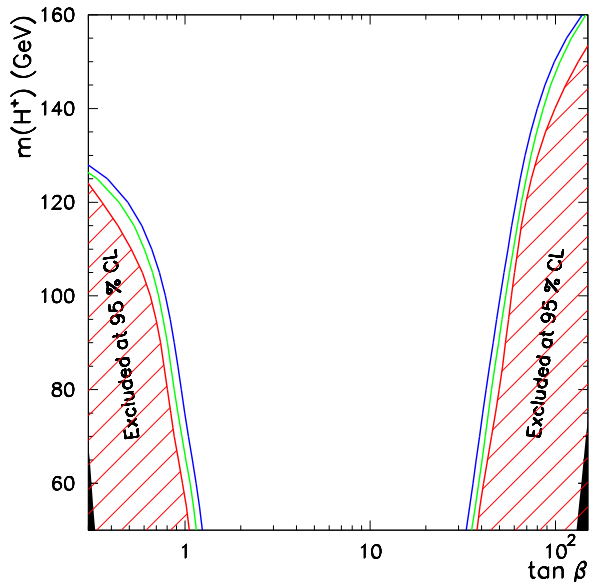}}\caption{D\O\ search for charged Higgs bosons in top
    decays \protect\cite{ehs-higgs-plus}. The hatched regions of scalar mass
    and $\tan\beta$ are excluded. }\label{fig6}}
\end{figure}

Many quite different kinds of models include relatively light charged scalar
bosons, into which top may decay: $t \to \phi^+ b$.  SUSY models must include
at least two Higgs doublets in order to provide mass to both the up and down
quarks, and therefore have a charged scalar in the low-energy spectrum.  The
general class of models that includes multiple Higgs bosons likewise often
includes charged scalars that could be light.  Dynamical symmetry
breaking models with more than the minimal two flavors of new fermions (e.g.
technicolor with more than one weak doublet of technifermions) typically
possess pseudoGoldstone boson states, some of which can couple to third
generation fermions.  Run I data already limits the properties of light
charged scalars coupled to t-b (see figure \ref{fig6}); Run II will explore the
remaining parameter space still further.  

\section{Low-scale top compositeness}
\setcounter{equation}{0} 

\begin{figure}[ht]
{\scalebox{.25}{\includegraphics[-4in,0in][6in,4in]{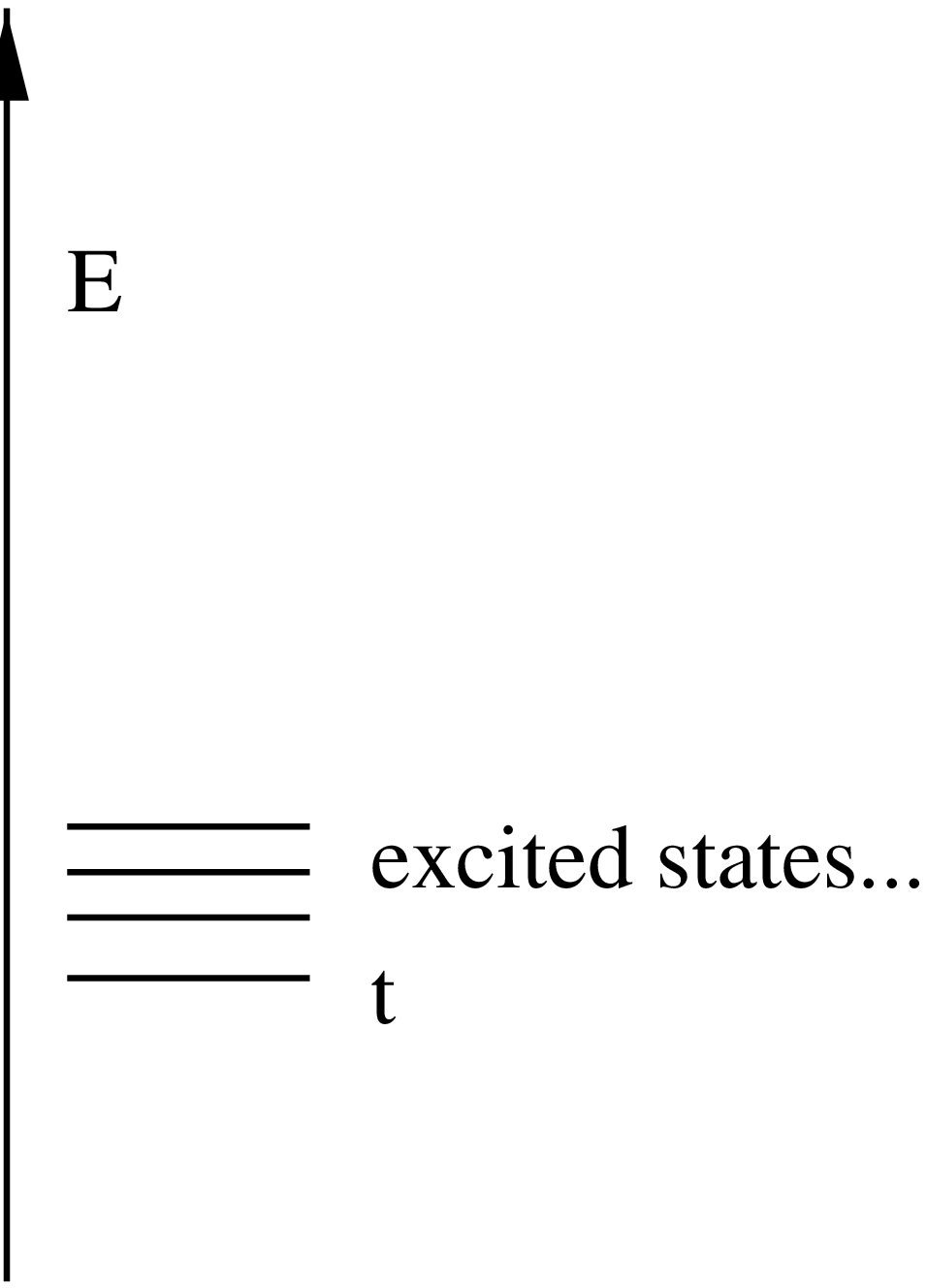}\hspace{1cm}
\includegraphics{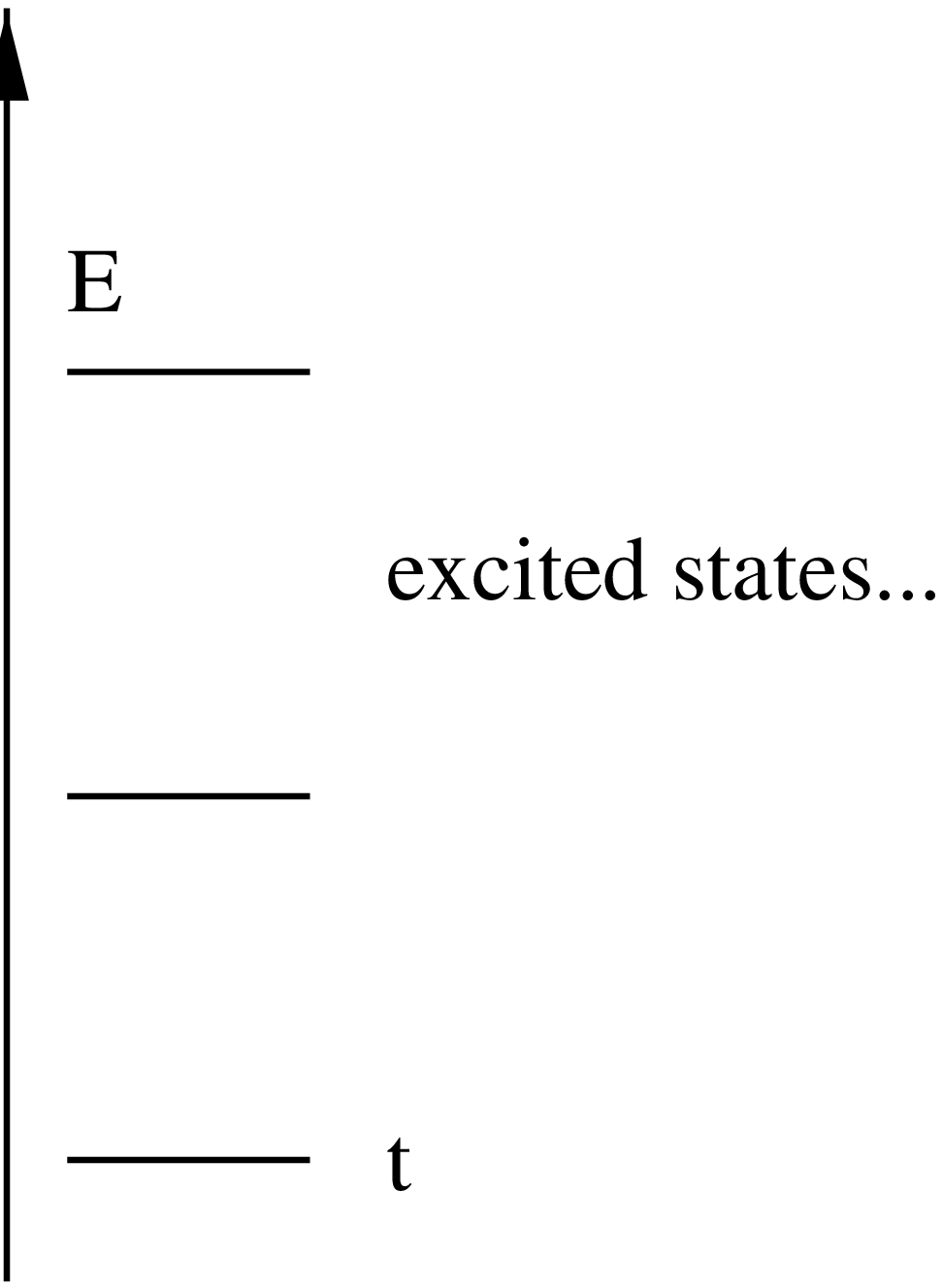}}\caption{A composite top quark would exhibit
  excited states.  Left:
      weak interactions underlying top compositeness produce inter-state
      spacing $\ll m_t$.  Right: strong interactions yield spacing $\gae
      m_t$.}\label{fig7}}
\end{figure}

We now turn to the possibility of a composite top quark.  Compositeness
requires new interactions to bind the consitutents together.  If those
interactions were weak, excited states of top would lie just above $m_t$;
strong coupling would produce large inter-state spacing (see figure
\ref{fig7}).  Since the three generations of quarks mix with one another, the
new interactions would couple at some level to first and second generation
quarks as well.  Thus, the absence of new weakly-coupled interactions of the
light fermions implies that top quark compositeness would have to arise from
strong interactions with a high intrinsic scale, $\Lambda$.

The magnitude of the effects of top compositeness on $q\bar{q} \to t \bar{t}$
depends on the properties of the constituents of the top.  
\begin{figure}[ht]{\scalebox{.25}{\includegraphics[-4.5in,0in][4in,3.75in]{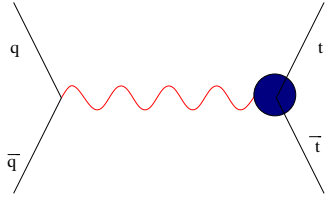}}
    \caption{  Composite top with colored constituents. $q\bar{q} \to t \bar{t}$
  scattering proceeds through gluon exchange:  $\sigma \approx \sigma_{SM} 
\left[ 1 + {\cal{O}}\left({\hat{s}\over\Lambda^2}\right)
\right]$}\label{fig8}}
\end{figure}
\noindent  If they carry color, scattering proceeds via gluon exchange and the 
cross-section is
modified from the QCD prediction by a form factor as in figure \ref{fig8}. This possibility and related effects like anomalous top chromomagnetic
moments have been studied in \cite{ehs-top-chrom}.  If the light quarks are
also composite and share
\begin{figure}[ht]
{\scalebox{.25}{\includegraphics[-4.5in,.25in][4in,4in]{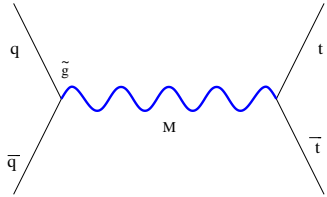}}
\caption{Composite  
      top and light quarks share constituents. $q\bar{q} \to t \bar{t}$
      scattering proceeds through interactions underlying compositeness: $\sigma
      \approx \sigma_{SM} \left[ 1 +
        {\cal{O}}\left({{\tilde{\alpha}\hat{s}}\over{\alpha_s M^2}}\right)
      \right] $}\label{fig9}}
\end{figure}
\noindent constituents with the top, scattering can
be caused directly by the interactions underlying compositeness (figure
\ref{fig9}) as well as by QCD gluon exchange.  As a result, the leading new
contributions to the scattering cross-section are enhanced by the strong
compositeness coupling, as envisaged in \cite{ehs-elp}.

\begin{figure}[hb]{\scalebox{.3}{\includegraphics[-4in,0in][4in,5.5in]{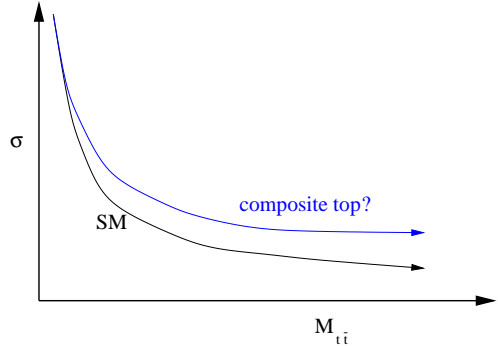}}\caption{Schematic
      invariant mass distribution of pair-produced top quarks in the standard
      model (SM) and assuming composite top quarks.}} \end{figure}

\section{Unusual quantum numbers}
 \setcounter{equation}{0}
 
 An idea that has received considerable attention recently is that the top
 quark could have gauge interactions differing from those of the other
 quarks.  Many of the proposed models involve extensions of one of the
 standard model gauge groups.  Specifically one of the $SU(N)$ groups in the
 standard model is assumed to arise from the spontaneous breaking of an
 $SU(N)_H \times SU(N)_L$ structure at higher energies, where some or all of
 the third-generation fermions transform under $SU(N)_H$ and all other
 fermions transform under $SU(N)_L$. The phenomenological result at low
 energies is typically the presence of a set of heavy vector bosons coupled
 primarily to $(t,b)$ or $(t,b,\nu_\tau,\tau)$.  Thus the low-energy
 properties of the first and second-generation fermions remain fairly
 standard, in agreement with experiment, while those of the third-generation
 fermions are modified in ways that may be visible.  

 In this section, we illustrate the possibilities, discuss their theoretical
 motivations, and note how one might test 
 them experimentally.  Section 6 shows a more complete model built using
 these principles.

\subsection{Extended Strong Interactions}
\setcounter{equation}{0}

One interesting possibility is to extend the strong interactions in a way
that causes them to distinguish among fermion flavors at energies 
above the weak scale.  At high energies, the strong interactions would then
include both an $SU(3)_H$ for the $t$ (and $b$) and an $SU(3)_L$ for the
other quarks.  to be consistent with low-energy hadronic data, these groups
must spontaneously break to their diagonal subgroup
(identified with $SU(3)_{QCD}$) at a scale $M$:
\begin{displaymath}
SU(3)_H \times SU(3)_L \to SU(3)_{QCD}\, .
\end{displaymath}
As a result of the symmetry breaking, a color octet of heavy gauge bosons
preferentially coupled to $t$ and $b$ is present in the spectrum at scales
below $M$.  

The extra gauge bosons have useful theoretical consequences.  Exchange of
\begin{figure}[hbt]
{\scalebox{.45}{\includegraphics[0in,1.9in][6in,7.5in]{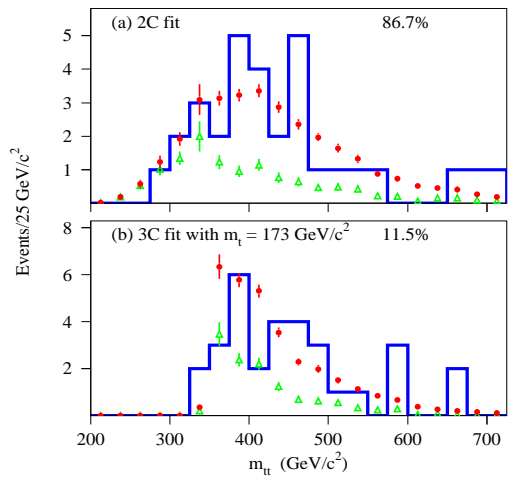}}
  \caption{Top-pair invariant mass spectrum from D\O\
    \protect\cite{ehs-watts}.  The 
    histogram 
    shows the data.  The open triangles are Monte Carlo background; the solid
    dots are MC background plus top signal.}\label{figd}} 
\end{figure}
the heavy gauge bosons yields a new four-fermion interaction
\begin{displaymath}
- {4\pi\kappa \over M^2}
\left(\overline{t}\gamma_\mu {\lambda^a\over 2} t\right)^2
\end{displaymath}
\begin{figure}[htb]
{\scalebox{.5}{\includegraphics[0in,3.5in][6in,9in]{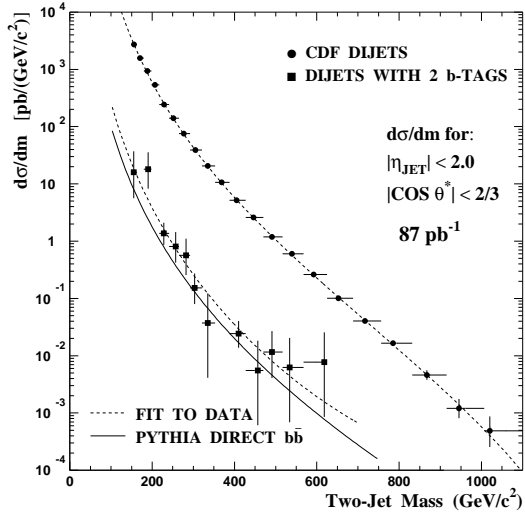}}
\caption{CDF dijet and $b\bar{b}$ spectra compared with PYTHIA standard model
    predictions \protect\cite{ehs-cdf-jets}.}\label{figc}} \end{figure}
\noindent that can cause top quark condensation ($ \langle \bar{t}t\rangle
\neq 0$) 
\cite{ehs-3-ttcond} .
This provides an opportunity for dynamical symmetry breaking to provide a
large mass for the top quark.  Furthermore, because the new interaction
treats top and bottom quarks identically, it need not make an unacceptably large
contribution to $\Delta\rho$.

Experimental tests of the extended strong interactions can be based on the
fact that the extra colored gauge bosons that become massive in this model
couple preferentially to the top and bottom quarks.  One may therefore, as
CDF \cite{ehs-cdf-jets} and D\O\ are already doing, seek evidence of new
resonances in the 
$t \bar{t}$ or $b\bar{b}$  invariant mass
spectrum (figures \ref{figd} and \ref{figc}) that do not also appear in 
the (light) dijet invariant mass spectrum.

\subsection{Extended Hypercharge Interactions}

A second possibility is to extend the hypercharge group to include a $U(1)_H$
felt by third-generation fermions and a $U(1)_
L$ felt by the light fermions.  Again, this extended group must be broken
at some high energy scale to its diagonal subgroup, which is identified with
the standard $U(1)_Y$:
\begin{displaymath}
U(1)_H \times U(1)_L \to U(1)_Y  \, .
\end{displaymath}
In the context of new strong dynamics, an extended hypercharge interaction
can be used to help generate the observed large splitting between the masses
of the top and bottom quarks, because these quarks carry
different values of hypercharge (see Section 6).

\begin{figure}[ht]
{\scalebox{.4}{\includegraphics[-2in,2in][5in,6.5in]{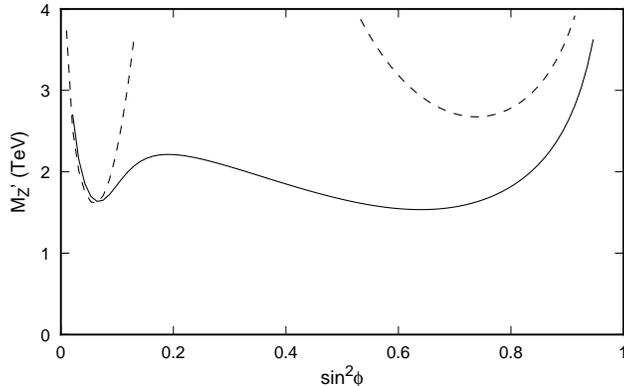}}
\caption{Lower bound on $Z'$ mass as a function of $Z$ $Z'$ 
  mixing angle $\phi$ \protect\cite{ehs-3-rscjtbd}.}\label{fig14}} 
\end{figure}

The broken hypercharge generator manifests itself physically as a heavy $Z'$
boson.  Indirect searches for such a $Z'$ look in precision low-energy and
$Z$-pole data for evidence of its mixing
with the ordinary $Z$.  A lower bound of 1.5 - 2 TeV on the
mass of the $Z'$ \cite{ehs-3-rscjtbd} has been set in this way (see figure
\ref{fig14}).  Direct
searches for a $Z'$ boson that couples preferentially to the third-generation
fermions can also be made in the invariant mass spectra of $t\bar{t}$,
$b\bar{b}$ and $\tau^+ \tau^-$. 

\subsection{Extended Weak Interactions}
Alternatively, one might extend the weak gauge group to include an $SU(2)_H$
felt by third-generation fermions and an $SU(2)_L$ coupled to the light
fermions \cite{ehs-2-ncetc}\cite{ehs-topfl}.  To preserve approximate weak
universality at low energies, this extended group must be spontaneously
broken at a high energy scale to its diagonal subgroup, which is identified
with the standard $SU(2)_W$:
\begin{displaymath}
SU(2)_H \times SU(2)_L \to SU(2)_W  \, .
\end{displaymath}
Because the breaking of the weak gauge group is central to generating
fermion masses, separation of the weak interactions of the heavy and light
fermions can allow distinct origins for their masses.  This can
help circumvent some of the traditional difficulties with constructing
dynamical models of mass generation.  

A class of dynamical models of this type \cite{ehs-2-ncetc}, called
``non-commuting extended technicolor'' (NC-ETC), has the symmetry-breaking
pattern
\begin{eqnarray*}
G_{ETC}  &\times& SU(2)_{L}\\
&\downarrow& \\
G_{TC} \times SU(&\!\!\!2&\!\!\!)_{H}  \times SU(2)_{L} \\
&\downarrow&\\
G_{TC}  &\times& SU(2)_{W}
\end{eqnarray*}
in which $SU(2)_H$ is embedded in the ETC interactions at high energies.
Cancellation between the effects of ETC gauge boson exchange and mixing
between the $Z$ bosons of the two $SU(2)$ groups enables $R_b$ to have a
value consistent with experiment.  At the same time, weak boson mixing causes
the weak interactions of the top quark to differ from those of the up and
charm quarks at low energies.

\begin{figure}[ht]
{\scalebox{.45}{\includegraphics[-1.5in,3.5in][7.5in,7in]
    {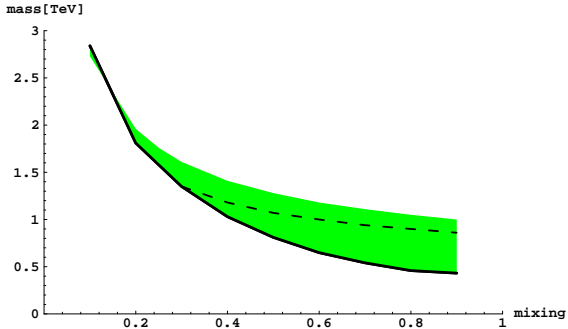}}\caption{The vertical axis is the heavy $W$'s mass; the
    horizontal axis is the degree of mixing of the two weak gauge groups in a
    non-commuting ETC model.  The area below the solid curve is excluded by
    precision electroweak data \protect\cite{ehs-2-ncetc}.  In the shaded
    region, $R_\sigma$ would be increased by at least 16\%
  \protect\cite{ehs-3-wtb2}. }\label{fig15}}
\end{figure}

Non-standard top quark weak interactions may be detectable in single
top-quark production at Run IIb \cite{ehs-3-wtb}\cite{ehs-3-wtb2}.  The ratio
of cross-sections $R_\sigma \equiv \sigma(\bar{p}p \to t b) / \sigma(\bar{p}p
\to l \nu)$ can be measured (and calculated) to an accuracy
\cite{ehs-3-wtbacc} of at least $\pm 8\%$.  In NC-ETC models, mixing of the
$W$ bosons from the two weak groups alters the light $W$'s coupling to the
final-state fermions, including top quarks.  So long as the heavy $W$ bosons
are not too massive, the result is a visible {\bf increase} in $R_\sigma$ (see
Figure \ref{fig15}).

\section{Unique Role in Electroweak Dynamics}
\setcounter{equation}{0} 

New physics associated with the top quark will
be most interesting if it helps explain electroweak symmetry breaking.  If
top squarks are discovered in the Run II ``top'' sample, one reason for
enthusiasm would be a first sighting of particles outside the standard model
spectrum; but even more important would be the proof that low-energy
supersymmetry must be included in any non-standard physics that seeks to
explain the origin of mass. If the reaction $t \to \phi^+ b$ is
observed in Run II, the immediate question will be ``Is $\phi^+$ a
Higgs or a technipion ?''.

In some theories, the top quark itself helps explain the origin of mass.
Those in which the top quark has new gauge interactions are of particular
interest, because they can help resolve some outstanding difficulties of the
original dynamical electroweak symmetry breaking scenarios.  A key challenge
for models of dynamical mass generation is to provide simultaneously
\newline \null \hspace{.5cm} $\bullet$ the correct $M_W$ and $M_Z$, with
$\Delta\rho \approx 0$ \cite{ehs-2-dirr}
\newline \null \hspace{.5cm} $\bullet$ both $m_t$ and $m_t - m_b$ large
\newline \null \hspace{.5cm} $\bullet$ $R_b$ near the standard model
value. \cite{ehs-2-sbs} 
\newline The original extended technicolor models have difficulty meeting
this challenge.  Dynamical models with extended weak interactions
\cite{ehs-2-ncetc} have more success, but no complete model has been
constructed.  Here, we focus on dynamical models with extended strong (and,
sometimes, hypercharge) interactions, known as ``topcolor-assisted
technicolor'', which have made progress on all three issues.


The prototypical topcolor-assisted technicolor model \cite{ehs-3-tc2}
has the following gauge group and symmetry-breaking pattern.
\begin{eqnarray*}
G_{TC} &\times& SU(2)_W \times \\
{U(1)_H \times U(1)_L} &\times& {SU(3)_H \times SU(3)_L}  \\
&\downarrow& \ \ M \gae 1 {\rm TeV} \\
G_{TC} \times SU(2)_{W} &\times&
  {U(1)_Y} \times {SU(3)_C} \\
&\downarrow&\ \ \ \Lambda_{TC}\sim 1 {\rm TeV} \\
{U(1)_{EM}} &\times& {SU(3)_{C}}\, .
\end{eqnarray*}
The groups $G_{TC}$ and $SU(2)_W$ are
ordinary technicolor and weak interactions; the strong and hypercharge
groups labeled ``H'' couple to 3rd-generation fermions and have stronger
couplings than the ``L'' groups coupling to light fermions
The separate $U(1)$ groups ensure that the bottom quark will not
condense when the top quark does.  Below the scale $M$, the Lagrangian
includes effective interactions for $t$ and $b$:
\begin{eqnarray} 
&-&{{4\pi \kappa_{tc}}\over{M^2}} \left[\overline{\psi}\gamma_\mu  
{{\lambda^a}\over{2}} \psi \right]^2 \\
&-&{{4\pi \kappa_1}\over{M^2}} \left[{1\over3}\overline{\psi_L}\gamma_\mu  
\psi_L + {4\over3}\overline{t_R}\gamma_\mu  t_R
-{2\over3}\overline{b_R}\gamma_\mu  b_R
\right]^2 \, . \nonumber
\end{eqnarray}
So long as the following relationship is satisfied (where the critical
value is $\kappa_c \approx 3\pi/8$ in the NJL approximation \cite{ehs-3-njl}) 
\begin{displaymath}
\kappa^t = \kappa_{tc} +{1\over3}\kappa_1 >
\kappa_c  >
\kappa_{tc} -{1\over 6}\kappa_1 = \kappa^b\, ,
\end{displaymath}
only the top quark will condense and become very massive \cite{ehs-3-tc2}.

The topcolor-assisted technicolor models combine the strong points of
topcolor and extended technicolor scenarios to give a more complete dynamical
picture of the origin of mass
features\cite{ehs-3-tc2phen}\cite{ehs-1-dynam}. Technicolor 
causes most of the electroweak
symmetry breaking, with the top condensate contributing a decay constant
$f \sim 60$ GeV; this prevents $\Delta\rho$ from being too large, as
mentioned earlier.  So long as the $U(1)_H$ charges of the
technifermions are isospin-symmetric, they cause no additional large
contributions to $\Delta\rho$.  Precision electroweak data constrain the mass
of the extra $Z$ 
boson in these models to weigh at least 1-2 TeV \cite{ehs-3-rscjtbd}.
ETC dynamics at a scale $M \gg 1$TeV
generates the light fermion masses and contributes about a GeV to the
heavy fermions' masses; this does not generate large corrections to
$R_b$.  Finally, the top condensate provides the bulk of the top quark
mass and the top-bottom splitting.  The unique role of the top quark is what
makes these models of mass generation viable.

\section{Conclusions}

Run I has already taught us some fascinating things about the top quark,
considered simply as a member of the Standard Model.  Run II will clearly
enable us to learn far more.  The quest for understanding electroweak
symmetry breaking and fermion masses points to physics beyond the Standard
Model.  This opens up the possibility that the top quark may have unusual
properties, some of which could become apparent during Run II.  Whether the
new physics associated with top is compositeness, new related states, new
gauge interactions, or something not yet imagined (!) it would be
tremendously exciting if it also helped reveal the origins of mass.

\bigskip \centerline{\bf Acknowledgments} 

Thanks are due to R.S. Chivukula for discussions on top compositeness.



\begin{thebibliography}{99} 
\frenchspacing 

\bibitem{ehs-disc-top} F. Abe et al., The CDF Collaboration, 
  \prl{74}{1995}{2626}; S. Abachi et al., The D\O\ Collaboration, 
  \prl{74}{1995}{2632}. 
\bibitem{ehs-willen} S. Willenbrock, ``Thinking About Top within the Standard
  Model,'' Presented at the Thinkshop on Top Quark Physics at Run II, 
  16-18 October, 1998, Fermilab, Batavia, IL, \hepph{9905498}.
\bibitem{ehs-1-dynam} For a review, see R.S. Chivukula, \hepph{9701322}.
\bibitem{ehs-1-susy} For a review, see S. Dawson, \hepph{9712464}.
\bibitem{ehs-1-stab} G. Anderson, D. Castano, and A. Riotto,
  \prd{D55}{1997}{2950} [\hepph{9609463}]; H. Murayama and M. Peskin,
  \arnps{46}{1996}{533} [\hepex{9606003}]. 
\bibitem{ehs-demina} R. Demina, ``Top Sample
   Checklist'', Thinkshop on top-quark physics for Run II, Fermilab, Batavia
 IL, October 16-18, 1998, http://b0nd10.fnal.gov/~regina/thinkshop/ts.html
\bibitem{ehs-3-slite} Steve Worm, CDF Collaboration, ``Tevatron searches for
  stops and sbottoms", SUSY99, Fermilab, IL, June 14-19, 
  1999. Transparancies are on the CDF web site, 
http://www-cdf.fnal.gov/physics/exotic/conference/conference.html .
\bibitem{ehs-3-stops} G.L. Kane and S. Mrenna, \prl{77}{1996}{3502}
  [\hepph{9605351}] ; 
G. Mahlon and G.L. Kane, \prd{55}{1997}{2779} [\hepph{9609210}];
  M. Hosch et al., \prd{58}{1998}{034002} [\hepph{97112324}]. 
\bibitem{ehs-mix-lite} E.H. Simmons, \npb{324}{1989}{315}; B.A. Dobrescu
  and C.T. Hill, \prl{81}{1998}{2634} [\hepph{9712319}].
\bibitem{ehs-higgs-plus} B. Abbott et al., D\O\ Collaboration 
  \prl{82}{1999}{4975} [\hepex{9902028}]. 
\bibitem{ehs-top-chrom} P. Cho and E. Simmons, 
  \plb{323}{1994}{401} [\hepph{9307345}]; D. Atwood, A. Kagan, and
  T.G. Rizzo,
  \prd{52}{1995}{6264} [\hepph{9407408}].
\bibitem{ehs-elp} E. Eichten, K. Lane, and M.E. Peskin, \prl{50}{1983}{811}
\bibitem{ehs-3-ttcond} V.A. Miransky, M. Tanabashi and K. Yamawaki, 
 \plb{221}{1989}{177}; \mpla{4}{1989}{1043}; Y. Nambu, EFI-89-08 (1989)
 unpublished; 
  W.J. Marciano, \prl{62}{1989}{2793}; 
  W.A. Bardeen, C.T. Hill and M. Lindner,  \prd{41}{1990}{1647}; 
    C.T. Hill, \plb{266}{1991}{419}.
\bibitem{ehs-watts} G. Watts, ``Top Quark: Experimental Status,'' 8th
  International Symposium on Heavy Flavour Physics, 
  Univ. of Southampton, Southampton, England, July 26-29, 1999.
\bibitem{ehs-cdf-jets} F. Abe et al., CDF Collaboration, \prl{82}{1999}{2038}.
\bibitem{ehs-3-rscjtbd} R.S. Chivukula and J. Terning, 
  \plb{385}{1996}{209},  [\hepph{9606233}].
\bibitem{ehs-2-ncetc} R.S. Chivukula, E.H. Simmons, and J. Terning, 
    \plb{331}{1994}{383} [\hepph{9404209}]; 
\prd{53}{1996}{258} [\hepph{9506427]}.
\bibitem{ehs-topfl} D.J. Muller and S. Nandi, 
\plb{383}{1996}{345} [\hepph{9602390}];  E. Malkawi, T. Tait, and C.-P. Yuan,
 \plb{385}{1996}{304} [\hepph{9603349}].
\bibitem{ehs-3-wtb} T. Stelzer and S. Willenbrock, \plb{357}{1995}{125} [\hepph{9505433}]. 
\bibitem{ehs-3-wtb2} E.H. Simmons, \prd{55}{1997}{5494} [\hepph{9612402}].
\bibitem{ehs-3-wtbacc} A.P. Heinson, \hepex{9605010}; A.P. Heinson,
  A.S. Belyaev and E.E. Boos, \hepph{9612424}; M.C. Smith and
  S. Willenbrock,  \prd{54}{1996}{6696} [\hepph{9604223}].
\bibitem{ehs-2-dirr} T. Appelquist et al., \prl{53}{1984}{1523} and
  \prd{31}{1985}{1676}.  
\bibitem{ehs-2-sbs} R.S. Chivukula, S.B. Selipsky and  E.H. Simmons, \prl{69}{1992}{575}, \hepph{9204214}.
\bibitem{ehs-3-tc2} C.T. Hill, \plb{345}{1995}{483} [\hepph{9411426}].
\bibitem{ehs-3-njl} Y. Nambu and G. Jona-Lasinio, \pr{122}{1961}{345}.
\bibitem{ehs-3-tc2phen}K. Lane and E. Eichten,  \plb{352}{1995}{382}
    [\hepph{9503433}]; R.S. Chivukula, B.A. Dobrescu, and J. Terning, 
    \plb{353}{1995}{289} [\hepph{9503203}]; G. Buchalla et al.,
    \prd{53}{1996}{5185} [\hepph{9510376}].  

\end{thebibliography}
\end{document}